\documentstyle[11pt,aaspp4]{article}

\begin{document}

\def\mathnew{\mathsurround=0pt}
\def\simov#1#2{\lower .5pt\vbox{\baselineskip0pt \lineskip-.5pt
        \ialign{$\mathnew#1\hfil##\hfil$\crcr#2\crcr\sim\crcr}}}
\def\simg{\mathrel{\mathpalette\simov >}}
\def\siml{\mathrel{\mathpalette\simov <}}
\def\Mesz{M\'esz\'aros~}
\def\beq{\begin{equation}}
\def\enq{\end{equation}}
\def\bea{\begin{eqnarray}}
\def\ena{\end{eqnarray}}
\def\bec{\begin{center}}
\def\enc{\end{center}}
\def\hl{\hline}
\def\tl{\tableline}
\def\fnum{F_{\nu_m}}
\def\half{ 1/2 }
\def\lbr{\linebreak\noindent}

\received{9/26/97}
\accepted{ }
\slugcomment{ApJ, subm 9/26/97}
%\journalid{XXX}{XXX 1997}
%\articleid{11}{14}
\lefthead{\Mesz, Rees \& Wijers}
\righthead{Sources of GRB Afterglow Diversity}
\tighten

\title{ VIEWING ANGLE AND ENVIRONMENT EFFECTS IN GRB: \\ 
        SOURCES OF AFTERGLOW DIVERSITY}

\author{P. \Mesz\altaffilmark{1}, M.J. Rees\altaffilmark{2} 
                  \& R.A.M.J. Wijers\altaffilmark{2} }

\altaffiltext{1}{Dpt. of Astronomy \& Astrophysics, Pennsylvania State 
University, University Park, PA 16803}

\altaffiltext{2}{Institute of Astronomy, Cambridge University, Madingley Road, 
Cambridge CB3 0HA, U.K.}

%\bec {\today} \enc

%\newpage

\begin{abstract}

We discuss the afterglows from the evolution of both spherical and 
anisotropic fireballs decelerating in an inhomogeneous external medium. We 
consider both the radiative and adiabatic evolution regimes, and analyze the 
physical conditions under which these regimes can be used. Afterglows may 
be expected to differ widely among themselves, depending on the angular 
anisotropy of the fireball and the properties of the environment. 
They may be entirely absent, or may be detected without a corresponding 
$\gamma$-ray event.  A tabulation of different representative light curves 
is presented, covering a wide range of behaviors that resemble what is 
currently observed in GRB 970228, GRB 970508 and other objects.

\end{abstract}


%\newpage

\section{Introduction}
\label{sec:intro}

The discovery of the afterglows of gamma-ray bursts (GRB) provides new
information which can constrain the models used to explain these objects.
Significant interest was aroused by the fact that several of the features 
reported in the first GRB detected over time scales $\simg$ days at X-ray (X) 
and optical (O) wavelengths, GRB 970228 (\cite{cos97a}) agreed quite well with 
theoretical expectations from the simplest relativistic fireball afterglow 
models published in advance of the observations (\cite{mr97a}; see also 
\cite{vie97a}). A number of theoretical papers were stimulated by this and
subsequent observations (e.g. \cite{tav97}; \cite{wax97a};\cite{rei97};
\cite{wrm97}, among others), and interest continued to grow as new 
observations provided controversial evidence for the distance scale and the 
possible host (\cite{sah97}). New evidence and new puzzles were added when 
the optical counterpart to the second discovered afterglow (GRB 970508) was 
attributed a cosmological redshift (\cite{metz97}), as well as a radio 
counterpart (\cite{fra97};\cite{tay97}) and X/O light curves showing a rise 
and decline (\cite{djorg97};\cite{fru97}). Some bursts, however, were detected 
only in X but not O (e.g. GRB 970828),
while some which would have been expected to be seen in X or O were not
(e.g.\ GRB 970111). 

Additional structure on the light curves has also emerged from a continued 
analysis of some of these objects down to the faintest flux levels.
The large variety of behaviors exhibited by afterglows, while clearly
compatible with relativistic fireball models, poses new challenges of
interpretation, e.g. \cite{wax97b};\cite{vie97b};\cite{kapi97};\cite{rho97};
\cite{pac97}. Some of the questions at the forefront of attention include
the effect of the external medium, the degree to which afterglows may be
considered to be isotropic events, and the effects of the radiative efficiency
on the evolution of the remnant. We address all three of these issues here.
We also clarify some of the issues that have been recently raised about
the dynamical effects of different radiative efficiency regimes. We then
discuss the possible variety of afterglow behavior that is expected from 
isotropic or anisotropic fireballs expanding in a medium which may be 
inhomogeneous, either due to external gradients or due to expansion in
an irregular cavity. We apply these models to interpret some of the salient 
observational features of several GRB afterglows, and discuss 
their possible use for predicting detection rates of X/O/R afterglows 
undetected in $\gamma$-rays, as well as some possible reasons for the 
non-detection of afterglows in GRB.

\section{Expansion Dynamics and Radiative Efficiency}
\label{sec:radef}

In some bursts (e.g. GRB 970508) the afterglow seems to contain a 
significant amount of energy compared to the typical (isotropic) estimate 
of $E\sim 10^{51}$ erg s$^{-1}$.  This led \cite{vie97b} to suggest that 
the afterglow must remain radiatively efficient $\sim$ weeks after the 
burst and evolve with $\Gamma \propto r^{-3}$. This regime was also 
considered in \cite{kapi97}, who refer to previous relativistic fireball 
models as radiating only a small fraction of the total kinetic energy, 
and go on to consider instantaneously cooling fireballs. It is important 
to discuss in more detail what are the conditions necessary for the
radiative efficiency having an effect on the dynamic evolution of an
expanding cloud.

The classical fireball models, with ``isotropic equivalent" energies of $E \sim
10^{51}$ erg and bulk Lorentz factors $\Gamma\sim 10^2-10^3$ have, in fact,
been generally been taken to be in the radiative stage during the $\gamma$-ray
event, i.e. radiative efficiency near unity, meaning that of order the initial total
kinetic energy of the protons is radiated in the observer-frame expansion time
(\cite{rm92}, and subsequent papers). However, for some parameters the bulk of
this energy can appear at energies other than MeV (\cite{mrp94};\cite{mr94}).
This high efficiency in the initial deceleration shock can occur if the electrons
are assumed {\it in the shock} to be heated to $\gamma_e \sim \xi_e (m_p/m_e)
\Gamma$ with $\xi_e$ reasonably close to unity. Experimental evidence from
interplanetary collisionless shocks indicates that this could be the case.
In such fireballs the electrons are likely to retain high radiative efficiency
for some time after the GRB, and energetically the most important are the
newly shocked electrons near the downwards evolving peak, of initial post-shock
energy $\gamma_e \sim \xi_e (m_p/m_e) \Gamma(t)$. This peak is generally where 
most of the electron internal energy is. In the regime where the peak electrons 
have high radiative efficiency, {\it if throughout the entire remnant volume the 
protons are able to establish (and remain in) equipartition with the electrons}, 
then the remnant evolves with $\Gamma \propto r^{-3}$ in a homogeneous external
medium (e.g. \cite{blamac76}; \cite{vie97a}; \cite{kapi97}). This follows simply 
from the momentum conservation law, if we can assume that the radiative losses
tap also the proton and magnetic energy (strong coupling), and the radiative 
time scale is shorter than the dynamic time scale, so that energy conservation 
cannot be used. This is the classical ``snowplow" approximation of supernova 
remnants. The alternative regime is that where the radiative losses do not
tap the proton and magnetic energy, only the electron energy (weak coupling),
and/or the radiative cooling time scale is longer than the dynamic time scale.
In this case, possibly after an initial short cooling of the electrons, one 
can assume energy conservation (most of the energy is in the protons and/or
magnetic fields), and one has $\Gamma \propto r^{-3/2}$ in a homogeneous 
medium (e.g. \cite{blamac76}, \cite{pacro93}, \cite{ka94b}, 
\cite{mr97a}). 

The (strong coupling) radiative regime and the adiabatic regime can be 
generalized to the case where the fireball moves into
an inhomogeneous external medium. We consider a spherical fireball of energy 
$E$ and bulk Lorentz factor $\Gamma$ which are independent of angle $\theta$,  
expanding into an external medium of density $n(r) \propto r^{-d}$ which is 
also independent of the angle, where $r$ is distance from the center of the 
explosion. The shocked gas evolves according to the conservation law
\beq
\Gamma^{1+A} r^3 n \propto \Gamma^{1+A} r^{3-d} \propto \hbox{constant}~,
\label{eq:isocons}
\enq
where $A=1(0)$ corresponds to energy(momentum) conservation, i.e. to the
$\hbox{adiabatic~(radiative)}$ regimes. These regimes must be understood
in a global or dynamic sense, as applying to the entire remnant, i.e.
baryons, magnetic fields, electrons, etc., or at any rate to its dynamically
dominant constituents. Since the observer-frame or detector time $t$ must 
satisfy $r \propto ct \Gamma^2$, we have
\beq
\Gamma \propto r^{-(3-d)/(1+A)} \propto t^{-(3-d)/(7+A -2d)}~~~,~~~
r \propto t^{(1+A)/(7+A-2d)}~.
\label{eq:isodyn}
\enq
If $d=0$ and a remnant starts out in the strong coupling radiative regime 
($A=0$) then $\Gamma \propto r^{-3}$. However after the expansion has 
proceeded for some time eventually the cooling time of the electrons at the 
peak of the distribution becomes longer than the expansion time, at which 
point energy conservation (adiabatic approximation) becomes valid 
(eq.\ref{eq:isocons} with $A=1$) leading to $\Gamma \propto r^{-3/2}$. 
% CH
The power law of equation (\ref{eq:isodyn}) is valid as long as the remnant is
relativistic, i.e., until the time $t_{nr}/t_o \sim \Gamma_o^{(7+A-2d)/(3-d)}$,
where $t_o \sim t_\gamma$ is the duration of the $\gamma$-ray burst itself and
$\Gamma_o$ is the initial Lorentz during the burst. For a homogeneous medium 
$d=0$ and $t_{nr}/t_o \sim \Gamma_o^{(7+A)/3}$, e.g. for an adiabatic remnant
$A=1$ with $\Gamma_o =10^3$ and $t_\gamma=1$ s, one has $t_{nr} \sim$ 1 year. 
The subsequent behavior in the nonrelativistic stage is described in \cite{wrm97}.
Shorter times for reaching the nonrelativistic stage are possible for a radiative 
remnant, or for a radiative stage followed by an adiabatic one, while longer times 
can occur for longer $t_\gamma$ or for expansion into a medium whose density 
decreases with distance ($0 \leq d \leq 3$).
% ECH

However, the conditions under which the remnant dynamics may be considered 
adiabatic or radiative is far from unambiguous, and is crucially dependent on 
poorly known questions about post-shock energy exchange between protons and 
electrons. Electrons cool quickly compared to protons, and it remains an unsolved 
question, of importance also in other areas of astrophysics, whether {\it behind the
shocks}, after the electrons have cooled, the protons remain hot (i.e. a two
temperature plasma, as in hot torus models of AGN), or whether they tend toward
some degree of equipartition with the cooled electrons by virtue of unknown fast
energy exchange mechanisms. The collisionless shock transition is the only place 
where one is guaranteed fast varying chaotic electric and magnetic fields which 
% CH
can lead to quick relaxation, and it is  known that an interplay between protons and 
magnetic fields can occur there which can lead to values close to equipartition.
Such equipartition is also inferred from measurements in the ISM. However, in 
fast evolving flows such as those in GRB, it is unclear whether such equipartition
can occur anywhere except possibly near the shock transition itself.
% ECH
If behind the shocks the protons are unable to quickly readjust to the 
electron losses, then even if the electrons are radiative ($a=1$) the protons can 
be adiabatic ($A=1$) leading to $\Gamma \propto r^{-3/2}$, in a homogeneous
external medium. A related problem is that, in order for the remnant to evolve
with $\Gamma \propto r^{-3}$ (again for a homogeneous external medium), the magnetic 
energy should also decrease, since if the latter were conserved it would soon 
dominate the total energy density and the remnant would evolve as a polytrope 
with adiabatic index 4/3 which leads to $\Gamma \propto r^{-3/2}$. Thus both the 
proton and the magnetic energy need to be transferred on a fast time scale to 
the electrons in order to ensure $\Gamma \propto r^{-3}$.
% CH
An additional factor that might contribute towards a steepening of the decay of 
$\Gamma$ are other energy losses, e.g. such as from the escape of accelerated
nonthermal protons from the shell, if these carry substantial energy. So far, there 
are neither detailed simulations nor experimental evidence concerning this in GRB.
In the absence of such losses, or of a quick energy exchange between post-shock 
protons plus magnetic fields and the electrons, one can therefore have a situation 
where the electrons are ``radiatively efficient", but the dynamics of the remnant 
expansion follows an ``adiabatic" law. This occurs if the electron cooling time is 
less than the expansion time. The shocked electrons can radiate up to half of the 
proton energy in the shocks, but the protons and the magnetic fields retain at 
least half and this would be enough to ensure a quasi-adiabatic dynamic evolution 
of the remnant with  $\Gamma \propto r^{-3/2}$. The latter is also true when the
electrons are adiabatic. For an inhomogeneous medium that decays with radius,
both of these decay laws would be flatter.
% ECH

There is no difficulty in treating the weak coupling case where the electrons
are radiatively efficient but the dynamics is adiabatic (\S \ref{sec:iso}). 
This regime is physically as plausible, if not more, as the strong coupling 
one where protons and fields exchange energy with electrons on a fast time scale.  
For reasons of simplicity, in \S\S \ref{sec:iso}, \ref{sec:ani} 
and in the rest of the paper we will assume that magnetic fields are near 
equipartition with the protons, which ensures a simple expression for the 
electron radiative efficiency in terms of only the synchrotron cooling time 
and the expansion time (the situation where inverse Compton (IC) cooling is 
important only introduces some extra changes in the way the synchrotron efficiency 
is defined). 

\section{ Spherical Inhomogeneous Models}
\label{sec:iso}  

As in the previous section, we consider a spherical fireball of energy $E$ 
and bulk Lorentz factor $\Gamma$ independent of angle $\theta$ expanding 
into an external medium of density $n(r) \propto r^{-d}$.
In the simplest afterglow model one considers the time evolution of the
radiation from the external medium shocked by the blast wave as it slows down. 
Denoting quantities in the comoving frame of the shocked fluid with primes,
in the post-shock region the density is $n'\sim n \Gamma$, the mean proton and 
electron random Lorentz factors are $\gamma_p \sim \Gamma$ and $\gamma_e \sim 
\xi_e (m_p/m_e)\Gamma$ (where $(m_e/m_p) \siml \xi_e \siml 1$ is the fraction
of the electron equipartition energy relative to protons), the turbulently 
generated magnetic field (assumed to build up to a fraction $\xi_B \leq 1$ of 
the field in equipartition with the random proton energy) is $B' \propto 
\xi_B n^{1/2} \Gamma$, and the peak 
of the electron synchrotron spectrum is at comoving frequency 
$\nu'_m \propto B' \gamma_e^2 \propto \xi_B \xi_e^2 n^{1/2} \Gamma^3 
\propto r^{-d/2} \Gamma^3$. The corresponding observer peak frequency is
\beq
\nu_m \propto \xi_B \xi_e^2 n^{1/2} \Gamma^4 
\propto t^{-[12-(d/2)(7-A)]/(7+A-2d)}
\enq
The synchrotron radiative efficiency at $\nu_m$ is  $e_{sy,m} 
\sim {t'}_{sy,m}^{-1}/({t'}_{sy,m}^{-1}+{t'}_{ex}^{-1}+{t'}_{other}^{-1})$ 
where  $t'_{sy,m}\propto 1/(\xi_e\xi_B^2\Gamma^3n)$ is 
the comoving synchrotron cooling time at $\nu_m$, $t'_{ex}$
is comoving expansion time or adiabatic cooling time and $t'_{other}$ is
any other loss mechanism, e.g. inverse Compton (IC), if important. 
In the limit where only synchrotron and/or  adiabatic losses are important
we may write $e_{sy,m} \sim (t'_{ex}/t'_{sy,m})^a $, which is unity in the
electron radiative ($a=0$) regime, and $\leq 1$ in the electron adiabatic 
($a=1$) regime. We have $e_{sy,m} \sim (\xi_B^2 \xi_e r n \Gamma^2)^a \propto
(\xi_B^2 \xi_e r^{1-d} \Gamma^2 )^a$. 

We first assume (\S \ref{sec:radef}) that protons and magnetic fields are 
strongly coupled to electrons, so if the electrons are radiative the entire 
remnant is radiative, and the index $A=a$. 
The comoving synchrotron intensity at the comoving peak frequency is
$I'_{\nu'_m} \propto n'_e (P'_{sy}/\nu'_m) c t'_{min} 
\sim n'_e (\gamma_e m_e c^2 /\nu'_m) c e_{sy,m} 
\propto \xi_B^{2a-1} \xi_e^{a-1} r^{a-d(a+\half)} \Gamma^{2a-1}$,
where $t'_{min} \sim t'_{sy} e_{sy,m}$ is the shortest of the possible cooling 
times (synchrotron or adiabatic, in the above approximation). The flux from
the relativistically expanding source at observer frequency $\nu_m$ is 
$F_{\nu_m} \propto t^2 \Gamma^5 I'_{\nu_m}$, or
\beq
\fnum
\propto \nu_m^{-[2(1-a)-(d/2)(1+a)]/[12-(d/2)(7-a)]} ~ 
\propto ~ t^{[2(1-a) -(d/2)(1+a)]/(7+a-2d)} ~,
\label{eq:fnisos}
\enq
scaling with $\xi_e^{a-1} \xi_B^{2a-1}$.
If the expansion is in the radiative $a=0$ regime, $\fnum \propto 
\nu_m^{-[2-(d/2)]/[12-(7/2)d]} \propto t^{(4-d)/(14-4d)}$ increases in 
time for any $d<3$, being $\propto \nu_m^{-1/6} \propto t^{2/7}$ in a 
homogeneous medium with $d=0$ (Vietri 1997b obtains a different scaling by
taking in $I'_{\nu_m}$ the shocked gas comoving width $\Delta R'$ as path
length, which however is equal to $c t'_{min}$ only for adiabatic expansion, 
e.g. \cite{mr97a}, where $t'_{min}\sim t'_{ex}$).
In the adiabatic $a=1$ regime $\fnum \propto \nu_m^{d/(12-3d)} \propto 
t^{-d/(8-2d)}$, which is a constant independent of $\nu_m$ and of time
for adiabatic expansion in a homogeneous $d=0$ medium 
(\cite{mr97a};\cite{ka94b}), but $\fnum$ decreases 
in time for an inhomogeneous medium with $0< d <3$ (for $d \geq 3$
the fireball encounters most of the external mass near its initial radius).
For a power-law spectrum $F_\nu \propto \nu^\alpha$, the flux at a fixed 
detector frequency $\nu_D$ is  
\beq
F_D = F_{\nu_m} (\nu_D /\nu_m)^{\alpha} ~\propto ~ 
 t^{[2(1-a)+12\alpha -(d/2)( \lbrace 1+a \rbrace +\alpha \lbrace 7-a \rbrace )]
 /[7+a-2d]} ~.
\label{eq:fdisos}
\enq
scaling as $\xi_e^{-1+a-2\alpha} \xi_B^{-(1+\alpha)+2a}$. Eq.\
(\ref{eq:fdisos}) is valid for the strong electron-proton coupling regime.
For a typical synchrotron spectrum $\alpha \simeq 1/3(-1)$ below(above) the 
break frequency $\nu_m(t)$, as the latter decreases in time the flux from 
radiative $a=0$ models in the detector frequency band $\nu_D$ at times for 
which $\nu_D < \nu_m$~,$\nu_D > \nu_m$ is $F_D \propto t^{6/7},~t^{-10/7}$ 
in a homogeneous $d=0$ medium, and $F_D \propto t^{8/9},~t^{4/3}$ in an 
inhomogeneous $d=2$ medium.  Adiabatic $a=1$ models give $F_D \propto 
t^{1/2},~t^{-3/2}$ in a homogeneous medium, and $F_D \propto t^0,~t^{-2}$ in
an inhomogeneous $d=2$ medium. Other values can be calculated from eq. 
(\ref{eq:fdisos}) for different $\alpha$ before and after the break. Tables 1 
and 2 give for the isotropic strong coupling case several examples in the
next to last column.

A different regime is obtained if one assumes that the protons and magnetic
fields are {\it not} strongly coupled to the electrons behind the shocks. In
this case the dynamics of the remnant as a whole is controlled by the index $A$
in eqs. (\ref{eq:isocons},\ref{eq:isodyn}), and as long as $\Gamma \gg 1$ the 
remnant is adiabatic with $A=1$ and $\Gamma \propto r^{-3/2}$ for a homogeneous
medium, whether the electrons are radiative or not, i.e. independent of $a$ 
(see \S \ref{sec:radef}). More generally, in this case $\Gamma \propto 
t^{-(3-d)/(8-2d)}$, $r \propto t^{2/(8-2d})$, and
\beq
\nu_m \propto \xi_B \xi_e^2 ~ t^{-(12-3d)/(8-2d)}~.
\label{eq:numisow}
\enq
In the synchrotron efficiency one must keep a separate index $a \neq A$ 
to account for the electrons being radiative or not. We have then
${I'}_{\nu_m} \propto \xi_B^{2a-1} \xi_e^{a-1} r^{a(1-d)-d/2} \Gamma^{2a-1}$,
and $\fnum \sim t^2 \Gamma^5 I'_{\nu_m}$ is
\beq
\fnum \propto \nu_m^{-[4(1-a)-d]/(12-3d)} ~\propto ~
    \xi_B^{2a-1} \xi_e^{a-1} ~ t^{[4(1-a)-d]/(8-2d)}~.
\label{eq:fnisow}
\enq
In a given fixed detector band $\nu_D$ one observes for a typical spectral 
shape $F_\nu \propto \nu^\alpha$ a time-dependent flux  
$F_D \propto \fnum \nu_m^{-\alpha}$, or
\beq
F_D \propto \xi_B^{(2a-1-\alpha)} \xi_e^{a-1-2\alpha} ~
   t^{[4(1-a)-d+\alpha(12-3d)]/(8-2d)}
\label{eq:fdisow}
\enq
%CH
This is valid in the relativistic expansion regime. After a remnant becomes 
nonrelativistic (see below equation 2), the flux in a homogeneous $d=0,~a=1$
medium would steepen to $F_D \propto t^{(3+15\alpha)/5} \propto t^{-12/5}$ for 
$\alpha=-1$ (\cite{wrm97}).
%ECH
The variety of time behaviors possible for relativistic isotropic models in both 
the strong and weak coupling cases depending on whether they are radiative or 
adiabatic is shown in the last two columns of Tables 1 and 2.

The time behavior given by eqs. (\ref{eq:fdisos},\ref{eq:fdisow}) can be 
complicated by at least two effects. One is that, unless observations start 
after the peak electrons are adiabatic, at some subsequent time the value of 
$a$ in eqs. (\ref{eq:fnisos},\ref{eq:fdisos},\ref{eq:fnisow},\ref{eq:fdisow}) 
switches from 0 to 1 as the peak electrons become adiabatic.  The second is 
that if the detector frequency $\nu_D > \nu_m$,the flow (being controlled by 
particles radiating at $\nu_m$) can already be adiabatic, while the smaller 
number of higher energy particles radiating at $\nu_D $ may still radiate 
efficiently. The frequency and time power law dependences of the flux before 
and after electron radiative inefficiency occurs are different.  For a steady
injection of accelerated electrons, the self-consistent electron energy
power law index $p$ in the presence of fast synchrotron losses is one power
steeper than the injected spectrum, and the self-consistent synchrotron power
law index $\alpha = (p-1)/2$ is a half power steeper than for the adiabatic
(negligible loss) case.  If the lowest energy electrons near the peak
$\gamma_e \sim \xi_e (m_p/m_e) \Gamma$ are radiatively efficient, all
electrons above that are as well. As the remnant evolves, the first electrons
to become inefficient are the lowest energy ones (in the peak corresponding to
$\nu_m(t)$), and a flattening break by 1/2 power in the photon spectrum at
frequency $\nu_b(t) > \nu_m(t)$ moves to frequencies increasingly higher than
$\nu_m(t)$. The electron Lorentz factor at which the synchrotron time just equals
the expansion time $r/c\Gamma$ is $\gamma_b \propto r^{[2-A(1-d)]/(1+A)}$
and the corresponding ``adiabatic" photon frequency is
\beq
\nu_b \propto B' \gamma_b^2 \Gamma \propto n^{1/2}\Gamma^2 \gamma_b^2
\propto r^{-[2-(3/2)d]} \propto t^{-[2-(3/2)d](1+A)/(7+A-d)}~,
\label{eq:nub} 
\enq
which can either decrease in
time for $d < 4/3$ (including a homogeneous medium with $d=0$) or increase
for $d > 4/3$ (although it always increases with respect to $\nu_m(t)$). For an
external medium whose density drops with radius faster than $d \simg 4/3$,
if initially $\nu_b > \nu_D$ it will always remain so, and the spectral index
remains adiabatic without change (until a much higher cutoff is reached where
the  acceleration becomes inefficient and the spectrum drops off
exponentially).
However for a homogeneous medium or one with $d<4/3$, if initially $\nu_b >
\nu_D$ the photon spectral index will at some later time steepen by 1/2 as
$\nu_b$ sweeps through the observing band $\nu_D$ and the observed spectrum
transitions from the adiabatic to the radiative regime.

\section{ Anisotropic Inhomogeneous Models}
\label{sec:ani}

The observed afterglow temporal decays are conventionally fitted by power-laws, and it is interesting to explore how the decay slopes would depend on the 
angular dependence of the dynamically relevant quantities of a fireball. 
To that effect, we consider anisotropic relativistic outflows where both the 
energy per unit solid angle and the bulk Lorentz factor depend on the angle 
$\theta$ as power-laws (at least over some range of angles), and also consider
the external density distribution to depend on radius as a power-law,
\beq
E \propto \theta^{-j}~~~,~~~\Gamma\propto\theta^{-k}~~~,~~~n\propto r^{-d}~.
\label{eq:anidep}
\enq
If there is a well defined jet, the normalizations of $E$ and $\Gamma$ may 
be different for material inside and outside the jet opening angle $\theta_o$.
At each angle the outflow starts converting a significant fraction of its
bulk kinetic energy into radiation when an external blast wave develops
at the angle-dependent deceleration radius $r_d \propto [E/n(r_d)\Gamma^2]^{1/3}$,
at an angle-dependent observer-frame (detector) time $t \sim r/ c \Gamma^2$. 
The $\theta$-dependence of eq.(\ref{eq:anidep}) implies that the deceleration 
blast wave at different angles occurs at
\bea
& \Gamma \propto t^{-k(3-d)/(8k-j-2dk)}~, &  r \propto t^{(2k-j)/(8k-j-2dk)} 
 \nonumber \\ 
& E \propto t^{-j(3-d)/(8k-j-2dk)}~, & \theta \propto t^{(3-d)/(8k-j-2dk)}~.
\label{eq:anidyn}
\ena
Depending on the normalization of (\ref{eq:anidep}) and causality 
considerations, the radiation from the blast waves occurring at increasing
$\theta$ at successive times $t$ can dominate the afterglow evolution (as 
opposed to the decay of $E$ and $\Gamma$ along the same $\theta$ as a 
function of time). For instance, if the event has been detected at 
$\gamma$-rays, and there is a jet of opening angle $\theta_o$, the observer 
is presumably within angles $\siml \theta_o$ from the axis. For $\Gamma= 
\Gamma_o (\theta/\theta_o)^{-k}$, in order for subsequent blast waves at $r=
r_d$ from $\theta > \theta_o$ to be observed at times $t > t_o$ one needs 
$\Gamma^{-1} \simg \theta$ to be satisfied, that is $\theta/\theta_o > 
(\theta_o \Gamma_o)^{1/(k-1)}$, or
\beq
t/t_o \simg (\theta_o \Gamma_o)^{(8k-j-2dk)/[(3-d)(k-1)]}~.
\enq
For values of $k<1$ the blast waves are detectable at all angles, but for 
$k > 1$ there are ranges of $k,j$ for which $\theta_o \Gamma_o$ is 
limited to values $\siml 5-10$ in order to detect the blast wave at 
reasonable $t/t_o$. In this case the initial part of the afterglow may
be due to the evolution in time of the gas responsible for the burst
initially observed, until such a time when the causality condition is 
satisfied for gas at larger angles, and the newly shocked gas at increasing
angles can become dominant in providing the observed flux. This second case 
introduces additional complexities and will not be discussed here, since 
even the simpler case first mentioned above will serve to illustrate the
point that a great variety can be expected in the temporal behavior of
afterglows.

For the conditions where the afterglow is dominated by the newly shocked
gas at increasing angles, the
observer-frame peak frequency of the synchrotron radiation spectrum from 
the blast waves (\ref{eq:anidyn}) coming from increasingly larger $\theta$
at increasing times $t$ is
\beq
\nu_m \propto \xi_B \xi_e^2 ~ n^{1/2} \Gamma^4 ~\propto ~
  t^{-[12k-d(3k+ j/2)]/[8k-j-2dk]}~.
\enq
The observer-frame intensity at this peak frequency is  $I_{\nu_m} =
\Gamma^3 I'_{\nu'_m} \sim \Gamma^2 (E e_{sy,m} /4\pi r^2 t \nu_m ) \propto
\xi_B^{-1} \xi_e^{-2} e_{sy,m}~t^{-d(k-j/2)/(8k-j-2dk)}$, where as in \S\ref{sec:iso} 
the synchrotron efficiency is $e_{sy,m} \sim (t'_{ex}/t'_{sy,m})^a \propto
(\xi_B^2 \xi_e r^{1-d} \Gamma^2)^a$ if synchrotron and adiabatic cooling 
are the two most important energy loss mechanisms (or its generalization 
if IC or other effects need to be included). We have then
$e_{sy,m} \propto t^{-a(4k+j(1-d))/(8k-j-2dk)}$, where $a=1(0)$ if the
peak electrons in the deceleration blast wave at the angle corresponding to 
detector time $t$ are adiabatic (radiative). (In this model the dominant
radiation is produced at the initial deceleration blast wave for that 
$\theta$, so $a$ does not enter in the dynamics, only in the radiative 
efficiency of the initial blast wave). The flux observed at $\nu_m$ from
the deceleration blasts at increasing $\theta$ is $\fnum \sim t^2 \Gamma^2 
I_{\nu_m} $, or
\bea
\fnum ~ \propto ~
 & \nu_m^{-[4k-2j-d(k-j/2)-a(4k+j \lbrace 1-d \rbrace )]/[12k-d(3k+j/2)] } 
     \nonumber\\
 \propto ~ & t^{[4k-2j-d(k-j/2) -a(4k+j \lbrace 1-d \rbrace )]/[8k-j-2dk]} ~,
\label{eq:fnanis}
\ena  
which scales with $\xi_B^{2a-1} \xi_e^{a-1}$. 
Depending on the normalization of eqs (\ref{eq:anidep}), the flux 
(\ref{eq:fnanis}) from increasing $\theta$ values can dominate the flux 
given by equations (\ref{eq:fnisos}) or (\ref{eq:fnisow}). (In other cases, 
one can approximate the evolution as the superposition of isotropic blast 
waves from individual $\theta$, which could in some cases be dominated by 
that of the central jet region).
At a fixed detector frequency $\nu_D$, the observed flux corresponding to eq. 
(\ref{eq:fnanis}) is
\beq
F_D \sim F_{\nu_m} (\nu_D/\nu_m)^{\alpha} ~ \propto ~
 t^{[k(4+12\alpha )-2j-d \lbrace k(1+3\alpha)-(j/2)(1-\alpha) \rbrace
    -a\lbrace 4k+j(1-d)\rbrace ] /[8k-j-2dk] }~,
\label{eq:fdanis}
\enq
which scales with $\xi_B^{2a-1-\alpha } \xi_e^{a-1-2\alpha }$.
For characteristic spectra with $\alpha$ positive (negative) below (above)
the break, this leads to detected fluxes which initially rise in time, and
then decay. However, a variety of behaviors are possible, including some
where the flux after the break passes through the detector window continues to
grow at a slower rate, or saturates. Note that the scaling of eq. 
(\ref{eq:fnanis}) with $\nu_m$ allows both for $\fnum$ to decrease or to 
increase as $\nu_m$ decreases in time, both in the radiative and adiabatic 
cases, depending on the values of $j$ and $k$ which characterize the angular 
dependence of $E$ and $\Gamma$.

\section{Discussion}
\label{sec:disc}

% CH 
\noindent{\it 5.1~~ Dynamics, Cooling and Decay Law.- }
In the simplest model where the GRB and the afterglow both arise from
an external shock (case a1 of \cite{mr97a}), to zeroth order the GRB
$\gamma$-ray flux should lie near the backward extrapolation of the 
afterglow, provided the basic conditions have not changed and the same 
radiation mechanisms are  responsible for both. This is clearly a rough
approximation, since it is likely that the GRB is initially radiatively 
efficient, and becomes radiatively inefficient at some later stage.
The observations of GRB 970228, especially following the HST observations of 
September 1997 (Fruchter, et al, 1997) indicate an optical flux decaying 
as an approximately constant power law in time. Superposed on this overall
long-term behavior, there may be shorter timescale variations (e.g. Galama, 
et al, 1997), which could be either a difference in calibration, or a real 
wiggle in the decay.

At late times (e.g. six months in GRB 970228) the remnant is most likely adiabatic.
One question that can arise is whether a simple external shock afterglow 
model whose dynamics is manifestly ``adiabatic" can be radiatively efficient 
enough to produce the initial relatively high X-ray and optical afterglow 
luminosity. For an afterglow luminosity $L_{X,O} \siml L_\gamma$, as observed,
from our discussion in \S \ref{sec:radef} this is not a problem.
At each radius, the electrons can radiate up to half of the total newly shocked
proton energy randomized in the shock transition, as long as the electron 
cooling time is shorter than the expansion time. The electrons can be 
radiatively efficient even when the dynamics of the remnant (i.e. the shell 
of hot protons and magnetic fields behind the shock, which provide most of 
the mass and inertia) follows an ``adiabatic" law $\Gamma \propto r^{-3/2}$ for
a homogeneous medium. Arguments were presented \S \ref{sec:radef} why the latter 
behavior may be more likely than a faster evolution with $\Gamma \propto r^{-3}$, 
a conclusion also supported by comparison with observations relating to the size 
of the GRB 970228 remnant (\cite{wax97c}). Note also that, from equation 
(\ref{eq:isodyn}), for expansion in medium whose density decreases, $\Gamma$ could 
drop even more shallowly than the above for a homogeneous medium.
As shown by \cite{tav97};\cite{wax97a}; \cite{wrm97}; \cite{rei97} and others, 
in GRB 970228 the initial $\gamma$-ray flux and the overall X/O afterglow behavior 
are in good agreement with a simple external shock where the observations started 
after $\nu_m <\nu_D$, without any substantial changes of slope during the observed 
decay phase. However, as discussed below, there are various mechanisms capable of
producing changes in the decay slope, which could be responsible for some of the 
reported departures from a simple power law behavior. 

\noindent{\it 5.2~~Intensity Offsets Between Afterglows and Main Burst .-}
In the case of GRB 970228, a backward extrapolation of the afterglow flux shows 
some hints of undershooting the $\gamma$-ray flux. If this were real, a slight 
undershooting could be due to the GRB being radiative initially (this is 
expected especially during the initial deceleration shock), with the afterglow 
dynamics becoming adiabatic soon afterwards. This would lead to a flattening of 
the spectral slope (e.g. second line, last two columns of Table 2). 

Another possibility (\cite{mr97a}; \cite{wrm97}; \cite{kapi97}) is that the 
$\gamma$-rays could have originated in an internal shock.  Internal shocks leave 
essentially no afterglow, yet they should be followed (eventually) by external 
shocks. An internal shock leaves unused anywhere from $\sim 20\%$ to $\simg 90\%$ of 
the total kinetic energy (\cite{rm94}; \cite{koba97}). The leftover energy is 
liberated in the external shock, most of whose initial radiation can come out at 
GeV energies because initially inverse Compton (IC) losses dominate over 
synchrotron losses; the initial synchrotron MeV radiation would then be typically 
low below the BATSE threshold (\cite{mr94}), but later on it becomes dominant
over IC (\cite{wax97b}). Thus at later stages the afterglow from the synchrotron
peak can have a flux level whose back extrapolation might overshoot the $\gamma$-ray
flux. Since external shocks are generally smoother (at most 3-5 pulses, 
\cite{pan97}), while internal shocks may be very variable (\cite{rm94};\cite{koba97}),
a natural conclusion (also reached independently by Piran, 1997) is that since 
afterglows appear to arise from external shocks, a burst where the gamma-ray 
light curve (or at least the last gamma-ray pulse of the light curve) is 
relatively smooth has a better chance of leaving behind a visible afterglow 
at lower frequencies.

A reason for offsets may also be (\S \ref{sec:ani}) that the relativistic outflow 
has an angle dependence in either the energy, the bulk Lorentz factor, or both.
The GRB itself may be due, for instance, to a high $\Gamma$ ejecta which shocks first,
while the afterglow could be dominated by slower material ejected at larger angles
relative to the observer, which shock later and produce softer radiation, but 
which could carry a substantial or even larger fraction of the total energy. 
If detected this would generally be as an upward offset of the afterglow relative 
to the GRB, since otherwise the afterglow would be dominated by the evolution of the
same material which gave rise to the GRB.

\noindent{\it 5.3~~Rise and Decay, Late Rises, Bumps .- }
An initial afterglow flux rise followed by a decay is a direct consequence of 
the simplest afterglow model (a1), and is true also of any generic peaked 
spectrum from an expanding cloud where the peak energy decreases in time
faster than the peak flux. Estimates for expanding clouds were made by 
\cite{pacro93} and \cite{ka94b} based on simplified radiation models. 
In the more detailed model (a1) of \cite{mr97a}, if the slope $\alpha$
at frequencies below $\nu_m$ is positive, $F_\nu \propto \nu^\alpha$, the 
initial rise is $F_D \propto t^{3\alpha/2}$ for $\nu_D < \nu_m$ in a 
homogeneous medium (\cite{wrm97}). For an ``average" GRB spectrum, $\alpha \sim 0$ 
below the break (\cite{band93}), which implies $F_{\nu_D}$ initially constant 
(\cite{mr97a}).  However, $\alpha \sim 0$ is only the average value; there are 
many GRB with $\alpha >0$ below the break (this is also the case for an ideal 
synchrotron spectrum $\alpha \sim 1/3$ below the break, e.g. \cite{mrp94}), and 
in such cases one obtains an initial power law increase in $F_D$. After the
frequency $\nu_m$ of the peak has dropped below the observing frequency $\nu_D$,
if the spectrum above the peak is $F_\nu \propto \nu^\beta$ with $\beta <0$,
the flux observed at $\nu_D$ starts to decay $\propto t^{-3\beta/2}$. This 
occurs after a time $t_D/t_\gamma \sim (\nu_\gamma /\nu_D)^{2/3}$ (for an
adiabatic remnant $\Gamma \propto t^{-3/8}$), which for observations in the
R-band is $t_{opt} \sim 10^4 t_\gamma$. A maximum at 1.5 days such as in GRB 
970508 is therefore compatible with the observed $t_\gamma \sim 5\times 10^1$ s.

However, there are other mechanisms which can give rise to optical fluxes rising
at times which could be even later than the above. One is an anisotropic outflow 
(\S \ref{sec:ani}), where the GRB and X-rays come from material close to the axis 
oriented near the l.o.s., and further off-axis material with a slower $\Gamma$ 
starts to decelerate after 1.5 days, say, or it has been slowed down enough that 
its light cone includes the l.o.s. after 1.5 days. Since beaming does not change 
the slope, if the spectrum is a pure power law then its slope would remain constant.
While a constant spectral slope is the simplest assumption, it is not a necessary
one. It is conceivable, for instance, that the slope of accelerated electrons (and
consequently of the synchrotron spectrum) could change as the bulk Lorentz factor
and the shock strength changes, or it might depend on other effects associated 
with the angular anisotropy of the outflow. 
An interesting result is the indication that the optical light
curve of GRB 970508 may have been steady or even decreasing (\cite{ped97})
before the 1.5 day rise phase preceding the maximum. This may be simply 
explained in an anisotropic model such as described in \S \ref{sec:ani}
where the indices $j$ and $k$ change to give a light curve transition in
this sense, or even more simply, by a bimodal model where one has a central
jet associated with the $\gamma$-ray event, whose tail is just seen to
decay, followed by the emission of a slower $\Gamma$ outflow over much wider
angles outside the jet which is responsible for the main part of the afterglow.
Another possibility for a late rise in the optical would be if a lower $\Gamma$ 
shell catches up with the main shock front with a comparable energy after 
$t\sim 1.5$ days, so we see then the  emission from the onset of deceleration 
of this late shell. Very prolonged optical decays with a shallow power law 
are possible in anisotropic models such as in \S\ref{sec:ani}, e.g. Table 2, 
lines 6 or 7.

An alternative explanation for a late turn-on of an afterglow may be that the
GRB occurs inside a very low density cavity inflated by the pulsar activity of 
one of the neutron stars in the progenitor binary. The shock, at least over 
a range of directions, would not arise until the ejecta
hits the wall of the cavity, and this could take a time of order weeks,
the ensuing shock being spread over a dynamic time scale sufficiently short
to produce a large flux per unit time. A characteristic feature of pulsar 
cavities is that they are usually asymmetric and irregular in shape, often
being elongated due to the proper motion of the energizing source. One would
naturally expect a wide variety of time histories for the afterglows arising
from the impact of a (possibly anisotropic) ejecta upon an irregularly shaped
cavity wall whose dimension (depending on direction respect to the line
of sight) may vary considerably.

\noindent{\it 5.4~~Peak Flux Level Evolution.-}
In some observed afterglows the flux level $F_\nu$ at lower energies is, 
at least initially, significant relative to that of the maximum gamma-ray 
flux (even if $\nu F_\nu$ is smaller). One prediction of the simplest model
is that the maximum value of the afterglow flux in every band, $F_{\nu_m}$,
is a constant.  An interesting case is that of GRB 970508, where the ratio 
of maximum $F_X$ to maximum $F_\gamma$ is $\sim 1/2$, while the ratio of
$F_O$ to $F_\gamma$ is $\sim 1/10$. Considering the drastic simplifications
involved in the simplest model (homogeneous medium, constant equipartition
fields, etc), this order of magnitude agreement is perhaps encouraging.
However, $F_{\nu_m} \sim$ constant is clearly an approximation which need
not always hold for less simple models.   In Table 1, the top four lines 
of the last two columns show that for isotropic but inhomogeneous outflows
one can expect $F_{\nu_m}$ to either increase or decrease with $\nu_m$ as
the later decreases, while the previous columns of the top four lines in
the same Table show that either of these behaviors may also arise as a result
of an anisotropic outflow.  Thus, the ratio of $F_{\nu_m}$ at $\gamma$-rays 
and lower wavelengths could be either larger or smaller than one, simply on 
this basis. In a simple bimodal anisotropic model, one can simply have more or
less energy in the large angle slower outflow seen later than in the early
and harder axial outflow.

A related question is the observability of radio fluxes of order mJy around
$10^{10}$ Hz, as reported for GRB 970508 (\cite{fra97}). Radio fluxes 
of this magnitude arise naturally in a simple isotropic homogeneous 
fireball model after a week or so, since the self-absorption frequencies 
(overestimated by $10^2$ in \cite{mr97a}) are in the range of $10^{11}$ Hz 
initially and drop to $10^{10}$ Hz in about a week, and the flux level is 
well below the brightness temperature limits for incoherent synchrotron 
radiation. What is more interesting is the relatively large value of the 
radio flux ($\sim$ mJy) relative to the O flux ($\sim 50 \mu$Jy). In this
afterglow therefore $\fnum$ first decreases with $\nu_m$ between $\gamma$ and 
O energies, but then increases with decreasing $\nu_m$ between O and the R(adio)
band. The simplest explanation may be in terms of a jet-wind two component model. 
The decrease between $\gamma$ and O could be due to expansion of a jet into an 
inhomogeneous medium (e.g. Table 1, fourth line), and the increase between O and 
R could be due to a surrounding low $\Gamma$ wind at larger angles, which shocks 
at later times, as suggested in \cite{wrm97}. The slow wind need not have an 
angular dependence; a growth and decay of the flux is approximated also by 
behaviors represented in Table 2 in the last two columns for expansion into 
either a homogeneous or an inhomogeneous external medium. A wind with $\Gamma 
\sim 3-10$ can match the delayed emergence ($\sim$ week) of the radio from the 
wind blast wave.  The tables give only illustrative values for selected spectral and
density exponents, which can be easily changed to fit a particular observed
rate of growth and decay. Actually, the radio flux of GRB 970508 at $10^{10}$ 
Hz (Frail et al 1997) at first increased and then appears to flatten, except 
for decaying oscillations which could be due to scintillation 
(\cite{good97},\cite{wax97c}), followed by a slow decline. 

\noindent{\it 5.5~~Burstless Afterglows and Afterglowless Bursts.-}
An interesting consequence of anisotropic models (\cite{mr97b},\cite{rho97}) 
is that there could be a large fraction of detectable afterglows for which no
$\gamma$-ray event is detected. If the observer lies off-axis to the jet, 
then the detected ``afterglow" can be approximated by the isotropic model 
calculated for an $E(\theta_{obs}),~\Gamma(\theta_{obs})$ corresponding to 
the offset angle $\theta_{obs}$ of the observer to the jet axis. As $\Gamma$ 
drops after the deceleration shock, the causal angle includes an increasing 
amount of the solid angle towards the jet as well as towards the equator, and
depending on the values of $j$ and $k$ the observed flux would generally
decrease in time. However for some choices of $j,k$ an increase in the
light curve might be possible, depending on the normalization. It could be 
that gravitational energy is converted more efficiently into kinetic energy
of expansion at large angles, where the opacity is larger (c.f. \cite{pac97}).
After $\Gamma(\theta_{obs})$ has dropped to the point where the central jet 
portion $\theta \siml \theta_o$ is detectable, the late stages of the jet 
emission would become visible, at a later stage when it is bright only at 
wavelengths longer than $\gamma$-rays. If the jet contained substantially more 
energy than the off-axis regions so that it dominates the flux even after 
expanding for a longer time than the initially observed off-axis region, 
one would expect an additional increase or flattening of the light curve at 
this point. Details would be further complicated by contributions from the 
equator and the back side of an opposite jet, if $\theta_{obs} \simg \pi/4$.
The statistics of afterglows not detected in $\gamma$-rays can be calculated
from equations (\ref{eq:anidyn}, \ref{eq:fnanis}).

The converse question is why some bursts (e.g.\ GRB 970111) have been
detected in $\gamma$-rays but not in X or O, even though it was in the
field of view of Beppo-SAX, which would have been expected to detect it
if the X to $\gamma$-ray ratio had been comparable to GRB 970228
(a weak X-ray afterglow may have been detected, \cite{cost97b}).
One reason may be if the $\gamma$-ray emission is due to internal 
shocks (which leave essentially no afterglows, \cite{mr97a}), and the 
environment has a very low density, in which case the external shock can occur 
at much larger radii and over a much longer time scale than in usual afterglows
and the X-ray intensity is below threshold for triggering. 
This may be the case for GRB arising from compact binaries which are 
ejected from the host galaxy into an an external environment which is much less
dense that the ISM assumed for usual models.
Another possibility for an unusually low density environment, made up only of 
very high energy but extremely low density electrons, is if the GRB goes off
inside a pulsar cavity inflated by one of the neutron stars in the precursor 
binary. Such cavities can be as large as fractions of a parsec or more, giving 
rise to a deceleration shock months after the GRB with a consequently much 
lower brightness that could avoid triggering and detection. 

The lack of an afterglow in some bursts may also be due to occurrence
in an unusually high density environment (e,g, a star-forming region, or the 
inner kiloparsecs of a late type spiral, where failed supernova or hypernova 
progenitors may reside, e.g. \cite{pac97}). This could lead to a more rapid 
onset of the deceleration leading to the X-ray phase, and it would also
imply an increased neutral gas column density and optical depth in front of
the source.  A special case is that of GRB 970828,
where X rays have been observed, but no optical radiation down to faint
levels (Groot et~al. 1997). The presence of a significant column density
of absorbing material has been inferred from the low energy turnover of
the X-ray spectrum (\cite{mur97}), and the corresponding dust absorption
may in fact be sufficient to cause the absence of optical emission (Wijers
\& Paczy\'nski, private communications). The difference between the low
density and high density environments cases could be tested if future
observations of afterglows reveal a correlation with the degree of galaxy
clustering or with individual galaxies.

In conclusion, the absence of detected afterglows in many bursts is not 
surprising, while there may be detected afterglows also in some cases
where a corresponding gamma-ray burst has not been detected; and when 
afterglows are detected, a wide diversity of behaviors may be the rule, 
rather than the exception.
% ECH

\acknowledgements{This research has been supported by NASA NAG5-2857, 
the Royal Society and NATO CRG-931446. We are grateful to A. Panaitescu and
the referee for useful comments.} 

%\newpage

%\newpage

\begin{table*}
\begin{center}
\begin{tabular}{rrrrrrrrrrrr}
\tl \tl
 $a$ & $d$ & $\fnum, ~ \nu_m$ &  $j=\atop 3k$ & $j=\atop 2k$ & 
 $j=\atop k$ &$j=0\atop k=1$ & $j=\atop -k$ & $j=\atop -2k$ & $j=\atop -3k$ & 
 $\hbox{Isotr}\atop\hbox{A=a}$ & $\hbox{Isotr}\atop\hbox{A=1}$ \cr 
\tl
0 & 0 & $\fnum\propto\nu_m^q$ & 1/6 & 0 & -1/6 & -1/3 & -1/2 & -2/3 & 
                                                      -5/6 & -1/6 & -1/3 \cr

0 & 2 & $\fnum\propto\nu_m^q$ & 1/3 & 0 & -1/5 & -1/3 & -3/7 & -1/2 &
                                                     -5/9 & -1/5 & -1/3 \cr

1 & 0 & $\fnum\propto\nu_m^q$ & 3/4 & 1/2 & 1/4 & 0 & -1/4 & -1/2 & -3/4 &
                                                                  0 & 0 \cr

1 & 2 & $\fnum\propto\nu_m^q$ & 2/3 & 1/2 & 2/5 & 1/3 & 2/7 & 1/4 & 2/9 &
                                                               1/3 & 1/3 \cr
\tl

0 & 0 & $\nu_m\propto t^p$ & -12/5 & -2 & -12/7 & -3/2 & -4/3 & -6/5 &
                                                     -12/11 & -12/7 & -3/2 \cr

0 & 2 & $\nu_m\propto t^p$ & -3 & -2 & -5/3 & -3/2 & -7/5 & -4/3 &
                                                         -9/7 & -5/3 & -3/2 \cr

1 & 0 & $\nu_m\propto t^p$ & -12/5 & -2 & -12/7 & -3/2 & -4/3 & -6/5 &
                                                     -12/11 & -3/2 & -3/2 \cr

1 & 2 & $\nu_m\propto t^p$ & -3 & -2 & -5/3 & -3/2 & -7/5 & -4/3 &
                                                         -9/7 & -3/2 & -3/2 \cr
\tl
\end{tabular}

\tablenum{1}
\caption{ Exponents of the power law dependence of the synchrotron peak flux 
$\fnum$ as a function of the time-varying synchrotron peak $\nu_m$ (top), and 
exponents of the time dependence of $\nu_m$ on observer (detector) time $t$ 
(bottom).  The first column indicates whether the electrons are in the 
radiative ($a=0$) or adiabatic ($a=1$) regime, and the second column indicates
the value of the exponent of the external medium density dependence on radius 
$n \propto r^{-d}$, $d=0$ being homogeneous. Columns 4 through 11 give the
exponents for the anisotropic model $E \propto \theta^{-j}$, $\Gamma \propto
\theta^{-k}$ of \S \ref{sec:ani} and various values of $j$ and $k$. The last
two columns on the right gives the corresponding exponents for the isotropic 
models of \S \ref{sec:iso}, the first being for the strong electron-proton
coupling $a=A$ case, and the second for the weak coupling case $A=1$. 
For $A=0 (1)$ the remnant as a whole is dynamically radiative (adiabatic). }
\end{center}
\end{table*}

%\newpage

\begin{table*}
\begin{center}
\begin{tabular}{rrrrrrrrrrrr}
\tl \tl
  &  &    &  $F_D \propto t^w$  & &  &  &  &  & \cr 
\tl
 $a$ & $d$ & $\alpha$ & $j=\atop 3k$ & $j=\atop 2k$ & 
 $j=\atop k$ & $j=0 \atop k=1$ & $j=\atop -k$ & $j=\atop -2k$ & $j=\atop -3k$ 
  & $\hbox{Isotr}\atop\hbox{A=a}$ & $\hbox{Isotr}\atop\hbox{A=1}$ \cr
\tl
0 & 0 & 1/3 &  2/5 & 2/3 & 6/7 & 1 & 10/9 & 6/5 & 14/11 & 6/7 & 1 \cr

0 & 0 & -1  & -14/5 & -2 & -10/7 & -1 & -2/3 & -2/5 & -2/11 & -10/7 & -1 \cr

0 & 2 & 1/3 & 0 & 2/3 & 8/9 & 1 & 16/15 & 10/9 & 8/7 & 8/9 & 1 \cr

0 & 2 & -1  & -4 & -2 & -4/3 & -1 & -4/5 & -2/3 & -4/7 & -4/3 & -1 \cr 

1 & 0 & 1/3 & -1 & -1/3 & 1/7 & 1/2 & 7/9 & 1 & 13/11 & 1/2 & 1/2 \cr

1 & 0 & -1  & -21/5 & -3 & -15/7 & -3/2 & -1 & -3/5 & -3/11 & -3/2 & -3/2 \cr 

1 & 2 & 1/3 & -1 & -1/3 & -1/9 & 0 & 1/15 & 1/9 & 1/7 & 0 & 0 \cr 

1 & 2 & -1  & -5 & -3 & -7/3 & -2 & -9/5 & -5/3 & -11/7 & -2 & -2 \cr 
\tl
\end{tabular}

\tablenum{2}
\caption{ Exponents of the power law dependence of the synchrotron flux
in a given detector band $\nu_D$ as a function of observer (detector) time 
$t$.  The first column indicates whether the electrons are radiative ($a=0$) 
or adiabatic ($a=1$), the second column gives the exponent of the external 
medium density dependence on radius $n \propto r^{-d}$, $d=0$ being 
homogeneous, column 3 gives the value of the spectral index $\alpha$ of the 
spectrum, $F_\nu \propto \nu^\alpha$, which may be 1/3 below the peak $\nu_m$,
and -1 above (although these indices can vary around these representative 
values). As the peak $\nu_m$ passes down through the observation band $\nu_D$ 
the index $\alpha$ changes from the positive value left of the peak to the 
negative to the right of the peak.  Columns 5 through 12 give the time 
exponents of the $F_D \propto t^w$ in the detector band for the anisotropic 
model $E \propto \theta^{-j}$, $\Gamma \propto \theta^{-k}$ of \S 
\ref{sec:ani} and various values of $j$ and $k$. The last two columns on the 
right give the corresponding exponents for the isotropic models of \S 
\ref{sec:iso}, the first being for the strong coupling case $a=A$ and the 
second for the weak coupling $A=1$. 
For $A=0 (1)$ the remnant as a whole is dynamically radiative (adiabatic). }
\end{center}
\end{table*}

\end{document}